\documentclass[aps,prb,twocolumn,showpacs,superscriptaddress]{revtex4-1}
\usepackage{graphicx}
\usepackage{bm}
\usepackage{color,soul}

\begin{document}

\title{Antiferromagnetism in kagome $\alpha$-Cu$_{3}$Mg(OH)$_6$Br$_2$}
\author{Yuan Wei}
\affiliation{Beijing National Laboratory for Condensed Matter Physics, Institute of Physics, Chinese Academy of Sciences, Beijing 100190, China}
\affiliation{School of Physical Sciences, University of Chinese Academy of Sciences, Beijing 100190, China}
\author{Zili Feng}
\affiliation{Beijing National Laboratory for Condensed Matter Physics, Institute of Physics, Chinese Academy of Sciences, Beijing 100190, China}
\affiliation{School of Physical Sciences, University of Chinese Academy of Sciences, Beijing 100190, China}
\author{Clarina dela Cruz}
\affiliation{Neutron Scattering Division, Neutron Sciences Directorate, Oak Ridge National Laboratory, Oak Ridge, Tennessee 37831, USA}
\author{Wei Yi}
\affiliation{Semiconductor Device Materials Group, National Institute for Materials Science, 1-1 Namiki, Tsukuba, Ibaraki 305-0044, Japan}
\author{Zi Yang Meng}
\affiliation{Beijing National Laboratory for Condensed Matter Physics, Institute of Physics, Chinese Academy of Sciences, Beijing 100190, China}
\affiliation{CAS Center of Excellence in Topological Quantum Computation and School of Physical Sciences, University of Chinese Academy of Sciences, Beijing 100190, China}
\affiliation{Songshan Lake Materials Laboratory , Dongguan, Guangdong 523808, China}
\affiliation{Department of Physics, The University of Hong Kong, China}
\author{Jia-Wei Mei}
\affiliation{Shenzhen Institute for Quantum Science and Engineering, and Department of Physics, Southern University of Science and Technology, Shenzhen 518055, China}
\author{Youguo Shi}
\email{ygshi@iphy.ac.cn}
\affiliation{Beijing National Laboratory for Condensed Matter Physics, Institute of Physics, Chinese Academy of Sciences, Beijing 100190, China}
\affiliation{Center of Materials Science and Optoelectronics Engineering，University of Chinese Academy of Sciences, Beijing 100049, China}
\affiliation{Songshan Lake Materials Laboratory , Dongguan, Guangdong 523808, China}
\author{Shiliang Li}
\email{slli@iphy.ac.cn}
\affiliation{Beijing National Laboratory for Condensed Matter Physics, Institute of Physics, Chinese Academy of Sciences, Beijing 100190, China}
\affiliation{School of Physical Sciences, University of Chinese Academy of Sciences, Beijing 100190, China}
\affiliation{Songshan Lake Materials Laboratory , Dongguan, Guangdong 523808, China}

\begin{abstract}
The antiferromagnetism in $\alpha$-Cu$_3$Mg(OH)$_6$Br$_2$ was studied by magnetic-susceptibility, specific-heat and neutron-diffraction measurements. The crystal structure consists of Cu$^{2+}$ kagome layers with Mg$^{2+}$ ions occupying the centers of the hexagons, separated by Br$^{1-}$ ions. The magnetic system orders antiferromagnetically at 5.4 K with the magnetic moments aligned ferromagnetically within the kagome planes. The ordered moment is 0.94 $\mu_B$, suggesting little quantum and geometrical fluctuations. By comparing the magnetic and specific-heat properties with those of the haydeeite, we suggest that $\alpha$-Cu$_3$Mg(OH)$_6$Br$_2$ may be described by the two-dimensional spin-$1/2$ Heisenberg kagome model and is in the region of the ferromagnetic-order side of the phase diagram.
	
\end{abstract}



\maketitle

\section{introduction}

The two-dimensional (2D) spin-$1/2$ kagome model has been intensely studied in theories because its rich ground states \cite{SachdevS92,JiangHC08,JansonO08,YanS11,JiangHC12,MessioL12,PunkM14,BieriS15,IqbalY15,KumarK15,GongSS16,LiaoHJ17,MeiJW17,YCWang2017a, YCWang2017b}. For example, the $S$ = 1/2 Heisenberg antiferromagnetic kagome model (AFKM) can give rise to ferromagnetic (FM) order, different types of antiferromagnetic (AFM) orders and quantum spin liquids (QSLs) \cite{MessioL12}. The AFKM may be realized in many minerals of the atacamite group with the molecular formula as Cu$_{3}$$T$(OH)$_6$$X$$_2$, where $T$ and $X$ are the 3$d$ nonmagnetic transition-metal (Zn, Mg) and the halogen elements (F,Cl,Br), respectively. The most well-known material is the herbertsmithite Cu$_3$Zn(OH)$_6$Cl$_2$, which shows no magnetic order down to at least 30 mK and is suggested to be a QSL \cite{NormanMR16,ShoresMP05,BertF07,MendelsP07,HeltonJS07,HanTH12,FuM15,HanTH16}. The structure of $\gamma$-Cu$_3$Mg(OH)$_6$Cl$_2$ is very similar to that of the herbertsmithite and may also host a QSL state \cite{ColmanRH11}. Recently, new materials Cu$_3$Zn(OH)$_6$FBr and Cu$_3$Zn(OH)$_6$FCl with similar structure have been successfully synthesized and shown to possibly host the gapped $Z_2$ QSL ground states \cite{FengZL17,WenXG17,WeiY17,FengZL19}. 

In the above materials, the kagome layers formed by Cu$^{2+}$ ions are separated with each other by non-magnetic Zn or Mg ions, which may be treated as diamagnetic dilution of the three-dimensional (3D) pyrochlore-like lattice. These non-magnetic ions can also occupy the center of the hexagons in the kagome layers, as found in kapellasite, $\alpha$-Cu$_3$Zn(OH)$_6$Cl$_2$ \cite{ColmanRH08} and haydeeite, $\alpha$-Cu$_3$Mg(OH)$_6$Cl$_2$ \cite{ColmanRH10}. In this structure, the coupling between the kagome layers is through the weak interlayer O-H-Cl bonding, which should result in highly 2D magnetic properties. It had been suggested that the kapellasite may be a gapless spin liquid or noncoplanar coboc2-type AF order \cite{JansonO08,FakB12}, but later measurements found strong Cu/Zn site mixing that makes the AFKM inappropriate \cite{KermarrecE14}. The haydeeite is a rare example of the FM order in the AFKM with $T_c$ at 4.2 K \cite{BoldrinD15}. Measurements on the single-crystal haydeeite further revealed strong anisotropic behaviors between the in-plane and out-of-plane magnetic properties \cite{PuphalP18}. Although the magnetic structure has not been unambiguously solved, the ordered moment is less than 0.2 $\mu_B$, suggesting strong quantum fluctuations. Therefore, the haydeeite may be in the proximity to the quantum phase transition between the Heisenberg kagome FM order and the Heisenberg cuboc2 AF order \cite{BoldrinD15}. 

The idea for the compound of Cu$_3$Mg(OH)$_6$Br$_2$ comes from the substitution of interlayer Cu$^{2+}$ in the barlowite Cu$_4$(OH)$_6$FBr, which has perfect Cu$^{2+}$ kagome layers with Cu$^{2+}$ ions between them and is antiferromagnetically ordered at about 15 K \cite{HanTH14,JeschkeHO15,HanTH16b,FengZL18,TustainK18}. It has been theoretically suggested that Zn$^{2+}$ or Mg$^{2+}$ ions can replace the interlayer Cu$^{2+}$ in barlowite and thus dilute the AF order to give rise to a QSL state as in herbertsmithite \cite{LiuZ15,GuterdingD16}. While this proposal has been shown to succeed in Cu$_3$Zn(OH)$_6$FBr \cite{FengZL17,WenXG17,WeiY17,FengZL18,SmahaRW18}, no Mg-substituted barlowite has been reported. In this paper, we follow the same route in synthesizing the Zn-substituted barlowite to grow Mg-substituted barlowite. However, the final product is Cu$_3$Mg(OH)$_6$Br$_2$ because MgF$_2$ cannot be dissolved in water. Since it has the same crystal structure as kapellasite and haydeeite, we label it as  $\alpha$-Cu$_3$Mg(OH)$_6$Br$_2$. The system orders antiferromagnetically at about 5.4 K but the configuration of the moments within the kagome plane is FM with the ordered moment of 0.94 $\mu_B$. Our results suggest that $\alpha$-Cu$_3$Mg(OH)$_6$Br$_2$ is in the FM region of the phase diagram in the AFKM.

\begin{figure}[tbp]
\includegraphics[width=\columnwidth]{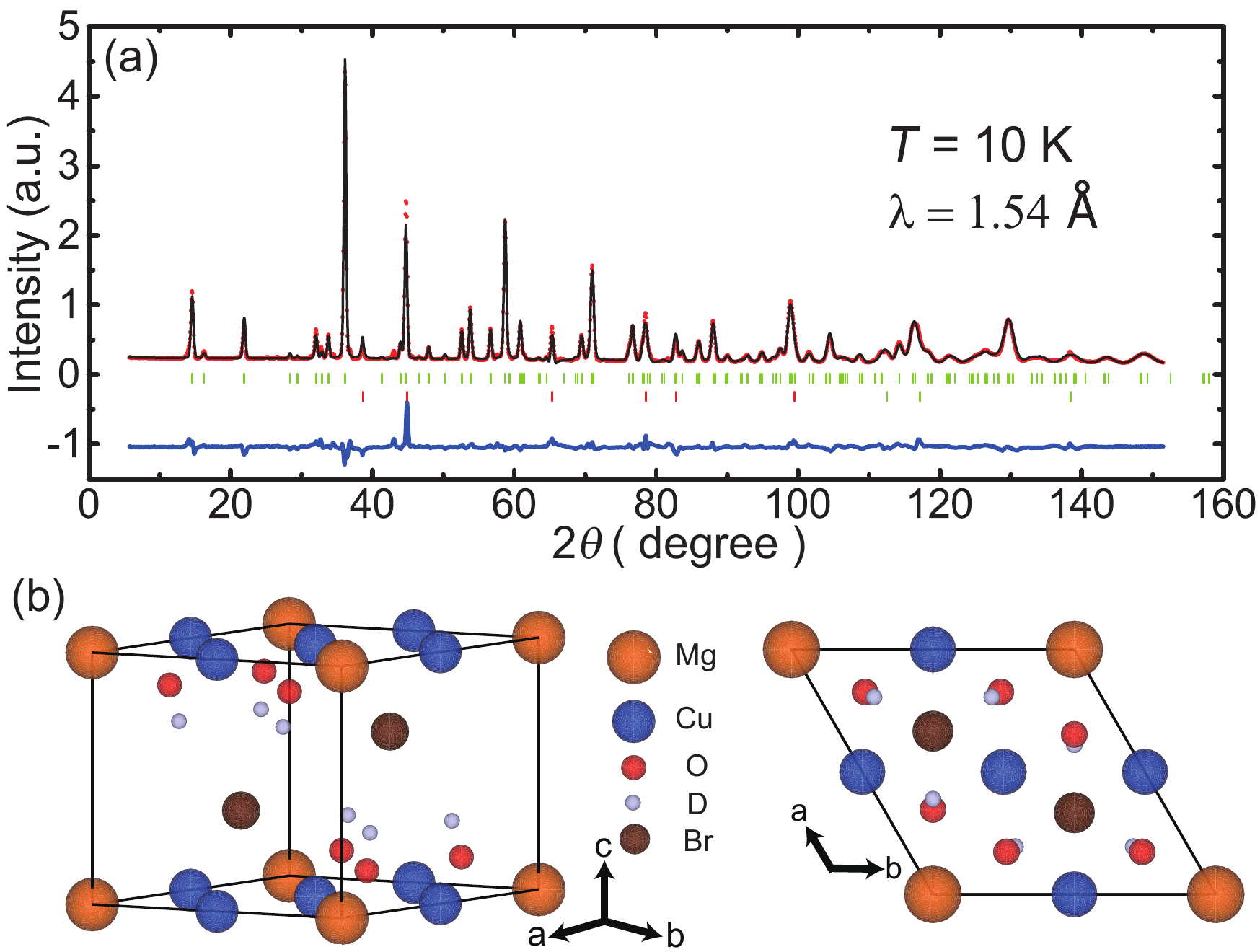}
\caption{(a) Neutron powder diffraction intensities of $\alpha$-Cu$_3$Mg(OD)$_6$Br$_2$ (red dots) at 10 K. The calculated intensities are shown by the black lines. Short vertical green lines represents Bragg peak positions, vertical red lines represents Al peaks. The blue line shows the difference between measured and calculated intensities. The weighted profile R-factor ($R_{wp}$) is 14.6\%. (b) Nuclear structure of $\alpha$-Cu$_3$Mg(OD)$_6$Br$_2$ (3D and top views).}
\end{figure}

\begin{table}
  \centering
    \begin{tabular}{ccccccc}
        \hline
          & Site & x & y & z & B (\AA$^2$)\\
        \hline
        Mg &    0.00000&  0.00000&  0.00000& 0.281(406)\\
        Cu &    0.50000&  0.00000&  0.00000& 0.131(131)\\
        Br &    0.33333&  0.66667&  0.63819(90)& 0.117(155)\\
        O  &    0.82592(32)&  0.17418(32)& -0.14100(64)&  0.511(120)\\
        D  &    0.19300(41)&  0.80700(41)&  0.28666(98)&  0.398(140)\\
        \hline
    \end{tabular}
  \caption{Nuclear structure parameters of $\alpha$-Cu$_{3}$Mg(OD)$_{6}$Br$_2$ at 10 K. $P6_{3}/mmc$ (No. 194): $a = b = 6.28014(20)$ \AA, $c = 6.06283(25)$\AA, $\alpha = \beta = 90 ^{\circ}$, $\gamma = 120 ^{\circ}$.}
\end{table}

\section{experiments}

$\alpha$-Cu$_3$Mg(OH)$_6$Br$_2$ was synthesized by the hydrothermal method as described previously \cite{FengZL17,FengZL18}. The mixture of 1.5-mmol Cu$_2$(OH)$_2$CO$_3$ and 6-mmol MgBr$_2$$\cdot$6H$_2$O was sealed in a 50-ml reaction vessel with 25-ml water, which was slowly heated to 200 $^\circ$C and kept for 12 hours. The polycrystalline samples were obtained by washing the production with the deionized water. To produce deuterated samples, Cu$_2$(OD)$_2$CO$_3$ and heavy water were used in the above process. The Mg content is determined by the inductively coupled plasma mass spectrometer. The magnetic susceptibility and heat capacity were measured by the MPMS and PPMS (Quantum Design), respectively. The magnetic and nuclear structures were determined by neutron diffraction experiments performed on the HB-2A diffractometer at HFIR, USA, with the wavelengths of both 2.4103 and 1.5395 \AA.

\section{results and discussions}

Figure 1(a) shows the neutron powder diffraction results at 10 K. There is no structural transition observed since the pattern at 300 K is similar to that at 10 K. Accordingly, the material has a hexagonal structure with the space group of $P6_{3}/mmc$. Detailed refinement results are shown in Table I. The large $R_{wp}$ is mainly due to the use of aluminum can and the presence of impurities. These chemical impurities only exist in the deuterated samples as shown by the room-temperature x-ray measurements. Figure 1(b) gives the nuclear structure. The Cu$^{2+}$ ions form kagome planes, which are separated by Br$^{1-}$ ions. The Mg$^{2+}$ ions sit at the centers of the hexagons within the Cu$^{2+}$ kagome planes. 

\begin{figure}[tbp]
\includegraphics[width=\columnwidth]{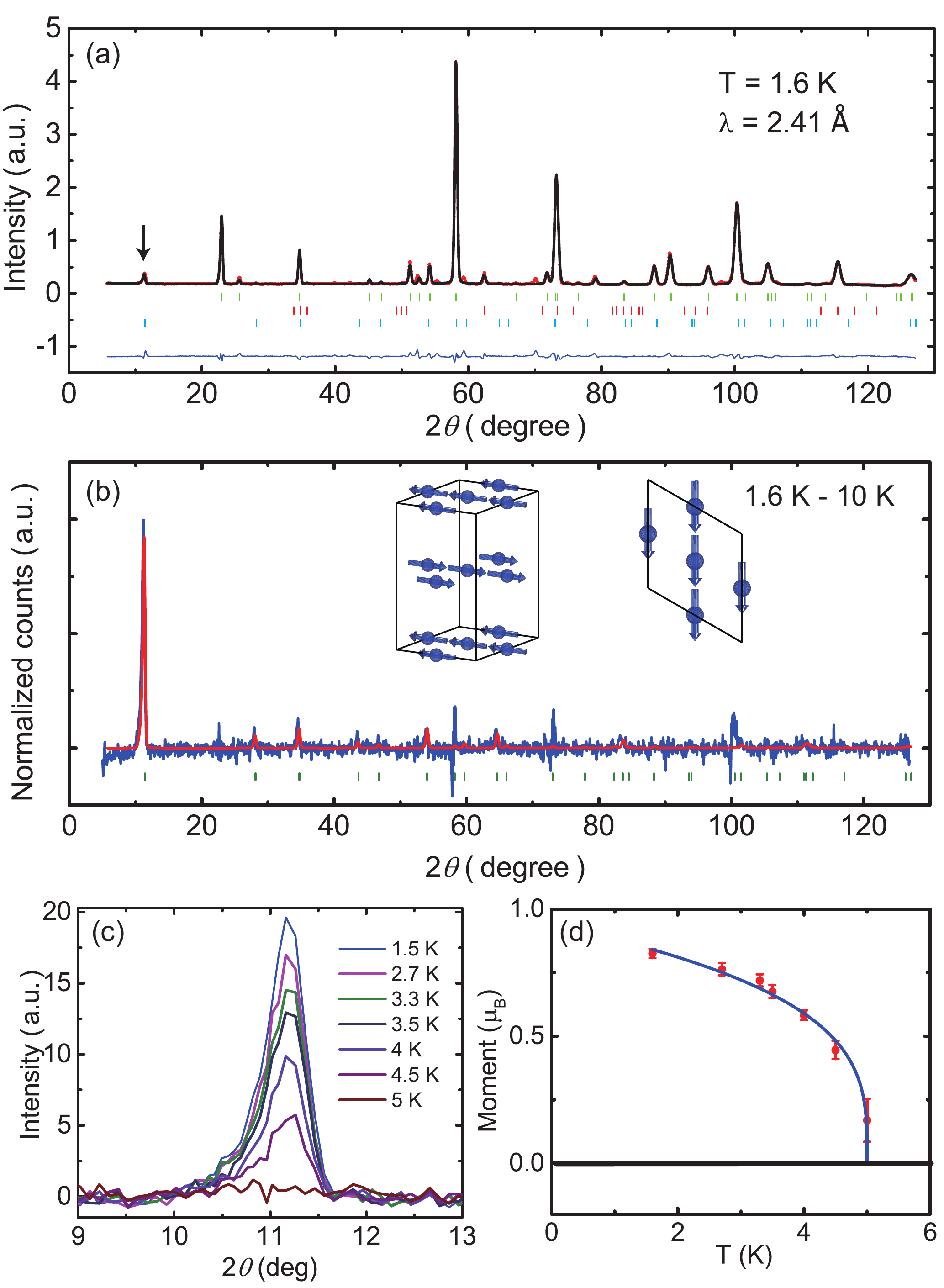}
\caption{(a) Neutron powder diffraction intensities of $\alpha$-Cu$_3$Mg(OD)$_6$Br$_2$ (red dots) at 1.6 K. The lines represent the same meanings as those in Fig. 1(a). The vertical blue lines represent magnetic peaks. The arrow shows the new peak appeared at this temperature. The weighted profile R-factor ($R_{wp}$) is 14.1\%.  (b) The difference (blue lines) between the intensities at 1.6 K and 10 K for $\alpha$-Cu$_3$Mg(OD)$_6$Br$_2$. The red lines represent the calculated intensities for the magnetic peaks. The vertical green lines represent magnetic peaks. The inset shows the magnetic structure. (c) The temperature dependence of the magnetic peak at (0, 0, 0.5). (d) The temperature dependence of the the moment. The error bars are from the refinements. The solid line is the fitted result as described in the main text with $T_N$ fixed at 5 K.}
\end{figure}

At 1.6 K, the system becomes magnetically ordered as shown by the new peak in Fig. 2(a). The subtraction between 1.6 and 10 K data provides more magnetic peaks as shown in Fig. 2(b). The k-search method is based on the first 5 possible magnetic peaks. The best result gives out a propagation vector $k$ = (0, 0, 0.5). Sarah was used to check the possible structures from the results of representational analysis based on the nuclear space group and magnetic k vector (0, 0, 0.5). There were 3 irreducible representations (IR or $\Gamma$ 1,3 and 5 ) as a result of the analysis. All the basis vectors were tried. The $\Gamma$ 1 and 3 can not fit the data at all. The best result comes from $\Gamma$5 BV 2. The other basis vectors for $\Gamma$5 other than Basis Vector 2 can not give good fits to the data. The inset of Fig. 2(b) shows the magnetic structure. The moments at Cu$^{2+}$ positions are confined within the kagome planes and ferromagnetically aligned along b-axis. Along the c-axis, the moments are aligned antiferromagnetically. 

Figure 2(c) shows the temperature dependence of the magnetic peak at (0, 0, 0.5) in the nuclear structural notation. Figure 2(d) shows the temperature dependence of the refined moment, which can be fitted by $M_0(1-T/T_N)^{\beta}$. The values of $M_0$ and $\beta$ are 0.94 $\pm$ 0.03 $\mu_B$ and 0.29 $\pm$ 0.03, respectively. The ordered moment $M_0$ at 0 K is consistent with the ordered moment $gS$ for a $S$ = 1/2 system with $g$ = 2. The value of $\beta$ is close to that in a classical 3D Ising (0.326) or Heisenberg (0.367) system.  

\begin{figure}[tbp]
\includegraphics[width=\columnwidth]{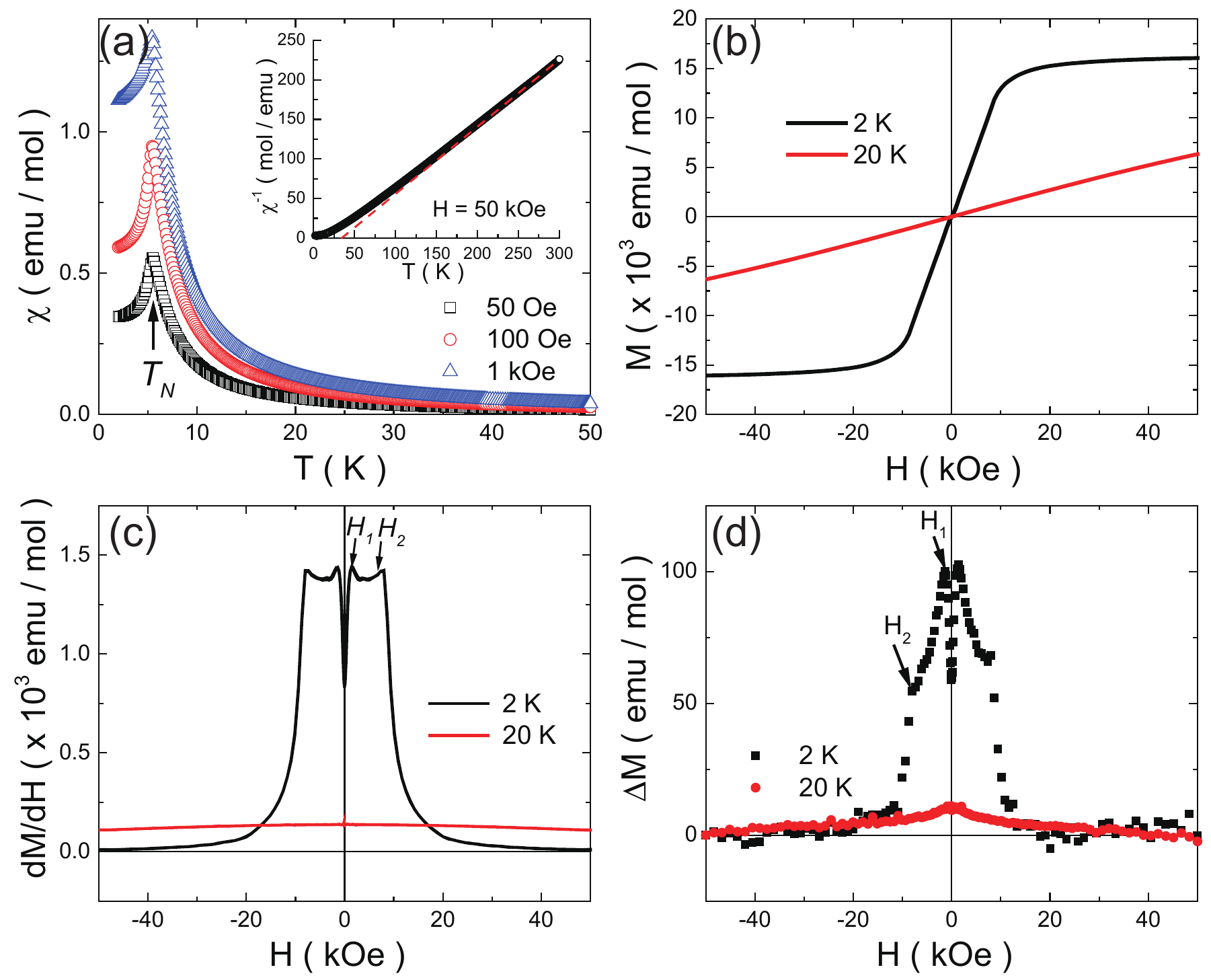}
\caption{(a) Temperature dependence of the magnetic susceptibility $\chi$ at 50 Oe, 100 Oe and 1 kOe. The arrow indicates the AF phase transition temperature. The inset shows the temperature dependence of $\chi^{-1}$ at 50 kOe. The dashed line is the linear fit to the high-temperature data. (b) $M$-$H$ loops at 2 and 20 K. (c) $dM/dH$ at 2 and 20 K. (d) The difference of $M$ between increasing and decreasing fields at 2 and 20 K.  }
\end{figure}

Figure 3(a) gives the temperature dependence of the magnetic susceptibility $\chi$, which clearly shows an AF phase transition at $T_N$ = 5.54 K. No significant difference is found between the field-cooled and zero-field-cooled processes. Above $T_N$, $\chi$ still strongly depends on the magnetic field, which suggests the presence of two-dimensional spin fluctuations. The inset shows the temperature dependence of $\chi^{-1}$ at 50 kOe. The fitting at high temperatures gives a Curie temperature $\theta$ of about 34 K and an effective moment of 1.77 $\mu_B$, which is close to the value for $S$ = 1/2 with g = 2. Figure 3(b) shows the field dependence of the magnetization $M$. At 2 K, $M$ becomes saturate above 20 kOe with the saturated moment of 0.95 $\mu_B$, which also suggests that the system is $S$ = 1/2. By taking the first derivative, two features can be seen at $H_1 \approx$ 1.4 kOe and $H_2 \approx$ 8 kOe as shown in Fig. 3(c). The value of $H_1$ is similar to the saturated field for field parallel to the kagome planes in haydeeite \cite{PuphalP18}, which suggests that the coupling between the kagome planes is weak. These two characteristic fields can also be seen by the hysteresis behavior as shown in Fig. 3(d). 

Figure 4(a) shows the specific heat of $\alpha$-Cu$_3$Mg(OH)$_6$Br$_2$ at several magnetic fields. The AF transition results in a large peak, which becomes a broad hump above 3 T due to the suppression of the transition. The background of the specific heat can be estimated by fitting the data from 15 to 30 K with $C = \alpha T^2 + \beta T^3$ as shown by the dashed line in Fig. 4(a). The fitted values for $\alpha$ and $\beta$ are 0.0265 J/mol K$^3$ and 2.86 $\times$ 10$^-4$ J/mol K$^4$, respectively, similar to those in barlowite \cite{HanTH14}. Figure 4(b) shows the temperature dependence of the entropy released associated with the magnetic transition, which is obtained by the integration of $C/T$ after subtracting the background as described above. The value of $\Delta S_{AF}$ at high temperature is about 3 J/mol Cu K, which is just about half of Rln2. This suggests that spin correlations are formed well above $T_N$, consistent with the results in the magnetic susceptibility measurements.

Interestingly, there is an upturn of $C$ below 0.2 K as shown in the inset of Fig. 4(a), which can be fitted as $C \propto T^{-3/2}$. This kind of temperature dependence suggests that the upturn is not from the nuclear schottky anomaly or magnetic impurities. Moreover, the specific heat should move to higher temperatures under fields if it comes from the nuclear schottky anomaly or magnetic impurities. Instead, it is suppressed at 30 kOe and thus should be related to the intrinsic kagome system before the spins are fully polarized by the field. The entropy between 0.07 K to 0.2 K is about 0.12 J/mol K. It is not clear what is the origin of this upturn, but it is rather surprising since the system shows no exotic properties from other measurements.

\begin{figure}[tbp]
\includegraphics[width=\columnwidth]{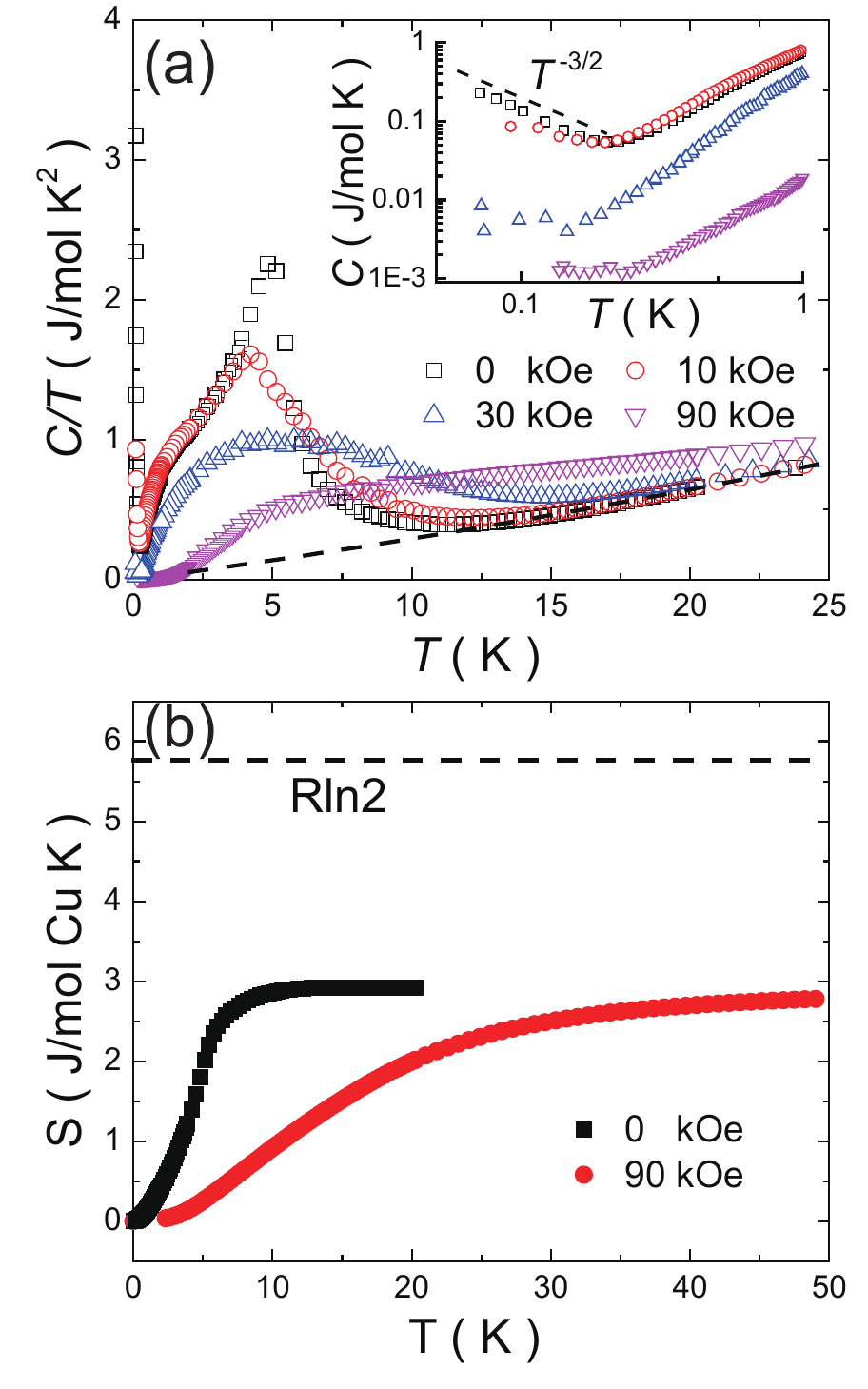}
\caption{(a) Temperature dependence of $C/T$ at several fields. The black dashed line is the fitted result for the phonon contribution as discussed in the main text. The inset shows the specific heat $C$ below 1 K. (b) Magnetic entropy associated with the magnetic transition at 0 and 90 kOe.}
\end{figure}

The above results suggest that $\alpha$-Cu$_3$Mg(OH)$_6$Br$_2$ exhibits 2D kagome ferromagnetism similar to haydeeite $\alpha$-Cu$_3$Mg(OH)$_6$Cl$_2$ \cite{BoldrinD15}, although the system orders antiferromagnetically along the c-axis. The ordered moment in haydeeite is just 0.2 $\mu_B$ due to strong quantum fluctuations, and it is suggested to sit near the boundary between the FM and cuboc2 type noncoplanar AFM parts in the phase diagram of the kagome lattice for a FM nearest-neighbour interaction \cite{BoldrinD15}. The $\alpha$-Cu$_3$Mg(OH)$_6$Br$_2$ is mostly likely in deep region of the FM part as its ordered moment is 0.94 $\mu_B$. While this makes $\alpha$-Cu$_3$Mg(OH)$_6$Br$_2$ less interesting due to the lack of frustration effects, it is worth noting that it may be another platform to study the topological bands in the kagome ferromagnet as observed in Cu[1,3-benzenedicarboxylate(bdc)] \cite{ChisnellR15}.

\section{conclusions}
Our systematically studies on the magnetism in $\alpha$-Cu$_3$Mg(OH)$_6$Br$_2$ demonstrate that it orders antiferromagnetically below 5.54 K but the spins ordered ferromagnetically within the kagome planes. The FM state of the kagome planes can be achieved easily by applying a weak magnetic field and its fluctuations survive above $T_N$. Our results suggest that $\alpha$-Cu$_3$Mg(OH)$_6$Br$_2$ is in the deep region of the FM part in the phase diagram of 2D AFKM. 

\begin{acknowledgments}
This work is supported by the Ministry of Science and Technology of China (Grants No. 2017YFA0302900, No. 2016YFA0300502, No. 2016YFA0300604), the National Natural Science Foundation of China (Grants No. 11874401, No. 11674406, No. 11574359, No.  11674370, No. 11774399, No. 11474330), the Strategic Priority Research Program(B) of the Chinese Academy of Sciences (Grants No. XDB25000000 and No. XDB07020000, No. XDB28000000), and the National Thousand-Young Talents Program of China. Research conducted at ORNL's High Flux Isotope Reactor was sponsored by the Scientific User Facilities Division, Office of Basic Energy Sciences, US Department of Energy.

Y.W. and Z.F. contributed equally to this work.
\end{acknowledgments}

\end{document}